\documentclass[11pt]{article}
\usepackage{fullpage}
\usepackage{cite}
\usepackage{graphicx}
\usepackage{amsmath}
\usepackage{amssymb}
\title{}
\date{}

\def\beq{\begin{equation}}
\def\eeq{\end{equation}}
\allowdisplaybreaks
\begin{document}
\bibliographystyle{utphys}
\newcommand{\msbar}{\ensuremath{\overline{\text{MS}}}}
\newcommand{\DIS}{\ensuremath{\text{DIS}}}
\newcommand{\abar}{\ensuremath{\bar{\alpha}_S}}
\newcommand{\bb}{\ensuremath{\bar{\beta}_0}}
\newcommand{\rc}{\ensuremath{r_{\text{cut}}}}
\newcommand{\Nd}{\ensuremath{N_{\text{d.o.f.}}}}
\setlength{\parindent}{0pt}

\titlepage
\begin{flushright}
QMUL-PH-17-14
\end{flushright}

\vspace*{0.5cm}

\begin{center}
{\bf \Large Extended solutions for the biadjoint scalar field}

\vspace*{1cm}
\textsc{Pieter-Jan De Smet\footnote{pejedees@yahoo.com}
and Chris D. White\footnote{christopher.white@qmul.ac.uk}} \\

\vspace*{0.5cm} Centre for Research in String Theory, School of
Physics and Astronomy, \\
Queen Mary University of London, 327 Mile End
Road, London E1 4NS, UK\\

\end{center}

\vspace*{0.5cm}

\begin{abstract}
Biadjoint scalar field theories are increasingly important in the
study of scattering amplitudes in various string and field
theories. Recently, some first exact nonperturbative solutions of
biadjoint scalar theory were presented, with a pure power-like form
corresponding to isolated monopole-like objects located at the origin
of space. In this paper, we find a novel family of extended solutions,
involving non-trivial form factors that partially screen the divergent
field at the origin. All previous solutions emerge as special cases.
\end{abstract}

\vspace*{0.5cm}

\section{Introduction}
\label{sec:intro}

The study of (quantum) field theories remains a highly active research
area, given that such theories describe the four fundamental forces in
nature. It is widely believed that these forces may emerge as the low
energy limit of an underlying framework, which may not necessarily be
a field theory (e.g. it may be a string theory). It is thus
interesting to elucidate common structures between theories, and also
to investigate how different types of theory are related to each
other. In this context, so-called {\it biadjoint scalar theory} has
recently been increasingly studied, whose Lagrangian is given by
\begin{equation}
{\cal L}=\frac12\partial^\mu\Phi^{aa'}\partial_\mu\Phi^{aa'}+\frac{y}{3}
f^{abc}\tilde{f}^{a'b'c'}\Phi^{aa'}\Phi^{bb'}\Phi^{cc'}.
\label{Lagrangian}
\end{equation}
Here the (un)primed indices are adjoint indices associated with two
different Lie groups, with structure constants $f^{abc}$ and
$\tilde{f}^{a'b'c'}$ respectively, and $y$ is a coupling
constant. Although this theory does not seem to be directly physically
applicable by itself, there is increasing evidence that it underlies
the dynamics in more relevant theories. For example, perturbative
scattering amplitudes in this theory can be used as building blocks
for amplitudes in non-abelian gauge
theories~\cite{BjerrumBohr:2012mg,Mafra:2016ltu}. The latter are
related to amplitudes in gravity theories by the {\it double copy} of
refs.~\cite{Bern:2008qj,Bern:2010ue,Bern:2010yg}, thus providing a
ladder of theories related by taking the field $\Phi^{aa'}$ of
eq.~(\ref{Lagrangian}), and replacing either of its adjoint indices
with Lorentz (spacetime) indices. This same procedure can be applied
to exact classical solutions as well as perturbative
amplitudes~\cite{Monteiro:2014cda,Luna:2015paa,Luna:2016due,Ridgway:2015fdl,Anastasiou:2014qba,Borsten:2015pla,Anastasiou:2016csv,Anastasiou:2017nsz},
where again the biadjoint theory plays a crucial role. Perturbative
radiative solutions have also been considered
recently~\cite{Goldberger:2016iau,Goldberger:2017frp,Luna:2016hge}, as
well as amplitudes in curved space~\cite{Adamo:2017nia}. A natural
framework for unifying the description of amplitudes in biadjoint,
gauge and gravity theories is the CHY equations of
refs.~\cite{Cachazo:2013iea,Cachazo:2013hca,Cachazo:2013gna,Cachazo:2013iaa},
which have been shown to emerge from a string theory in ambitwistor
space~\cite{Mason:2013sva,Geyer:2014fka,Casali:2015vta,Geyer:2015bja,Geyer:2015jch,Geyer:2016wjx},
itself a limit of conventional string
theory~\cite{Casali:2016atr}. Loop level aspects of biadjoint theories
in this formalism have been recently explored in
ref.~\cite{Gomez:2017cpe}. For related studies, see also
refs.~\cite{Chiodaroli:2016jqw,delaCruz:2016gnm,Cardoso:2016ngt,Mafra:2016mcc,Carrasco:2016ldy,Mizera:2016jhj,Campiglia:2017dpg,Johansson:2017srf}.\\

Given the wide range of instances in which the biadjoint scalar theory
appears, it is clearly worth studying this theory in its own
right. All of the above examples of its use involve perturbative
solutions of the equation of motion, which from eq.~(\ref{Lagrangian})
is found to be
\begin{equation}
\partial_{\mu}\partial^{\mu}\Phi^{aa'}-yf^{abc}\tilde{f}^{a'b'c'}\Phi^{bb'}\Phi^{cc'}=0.
\label{EOM}
\end{equation}
Instead, one may consider exact nonlinear solutions, which may involve
inverse powers of the coupling $y$. A number of solutions of this type
were recently presented in ref.~\cite{White:2016jzc}, consisting of
singular point-like disturbances located at the origin. The first of
these has spherical symmetry, and is applicable when the two Lie
groups are the same as each other:
\begin{equation}
\Phi^{aa'}=-\frac{2\delta^{aa'}}{y T_A r^2}.
\label{phisol1}
\end{equation} 
Here $r$ is the radial space coordinate~\footnote{We use the metric
  $(+,-,-,-)$ throughout.}, and we define
\begin{equation}
f^{abc}f^{a'bc}=T_A\delta^{aa'}.
\label{TAdef}
\end{equation}
Further solutions are possible if one restricts both Lie groups to
$SU(2)$, which allows mixing between spacetime and adjoint indices,
analogous to known non-perturbative solutions in nonabelian gauge
theories~\cite{Prasad:1975kr,Bogomolny:1975de,Julia:1975ff,'tHooft:1974qc,Polyakov:1974wq,Wu:1967vp}. In
particular, ref.~\cite{White:2016jzc} presented a one-parameter family
of solutions with an axial form in the internal space:
\begin{equation}
\Phi^{aa'}=\frac{1}{yr^2}\left[-k\left(\delta^{aa'}-\frac{x^a\,x^{a'}}
{r^2}\right)
\pm\sqrt{2k-k^2}\,\frac{\epsilon^{aa'd}x^d}{r}\right],
\label{phisol2}
\end{equation}
where $0\leq k\leq 2$ if $\Phi^{aa'}\in\mathbb{R}$. Like
eq.~(\ref{phisol1}), this has a pure power-like behaviour in
$r$. Indeed, this is the only possibility if an axial form is
imposed~\cite{White:2016jzc}, such that it tempting to think that the
spectrum of nonperturbative solutions in biadjoint scalar theory is
much simpler than that of non-abelian gauge theories. The aim of this
paper, however, is to show that extended solutions do in fact
exist. That is, it is possible to dress the power-like divergence in
$r$ with a non-trivial form factor, such that all previously found
solutions emerge as special cases. The new solutions are still
singular at the origin (a consequence of Derrick's theorem for scalar
field theories~\cite{Derrick:1964ww}, which prohibits finite
energy). However, one solution in particular will exhibit an
interesting screening behaviour, such that the strength of the
divergence in energy is ameliorated. Our results will be useful in
future studies of biadjoint scalar theory, including the issue of
whether or not double copy-like relationships can be generalised
beyond the perturbative sector.\\

The structure of the paper is as follows. In section~\ref{sec:common},
we present extended solutions for the case in which both Lie groups in
the biadjoint scalar Lagrangian are the same. In
section~\ref{sec:SU(2)}, we apply similar techniques to construct an
extended solution for the case in which both Lie groups are
$SU(2)$. We discuss our results and conclude in
section~\ref{sec:conclude}.

\section{Extended solutions for common Lie groups}
\label{sec:common}

In this section, we consider the Lagrangian of eq.~(\ref{Lagrangian}),
but where both sets of structure constants are associated with the
same Lie group. We may then make a similar ansatz to that made in
ref.~\cite{White:2016jzc}, namely
\begin{equation}
\Phi^{aa'}=\delta^{aa'}S(r), \quad S(r)=\frac{\bar{S}(r)}{y\, T_A}.
\label{ansatz1}
\end{equation}
Substituting this in eq.~(\ref{EOM}) yields a second-order non-linear
differential equation for $\bar{S}(r)$:
\begin{equation}
\frac{1}{r^2}\frac{d}{dr}\left(r^2\frac{d\bar{S}(r)}{dr}\right)
+\bar{S}^2(r)=0.
\label{Seq}
\end{equation}
We now look for extended solutions in which the $r^{-2}$ divergence of
the field at the origin, of eq.~(\ref{phisol1}), is dressed by a
finite form factor. To this end, we may write
\begin{equation}
K(r)=1+r^2\bar{S}(r),
\label{Kdef}
\end{equation}
where $K(r)$ is finite for all $r$. Equation~(\ref{Seq}) then becomes
\begin{equation}
r^2K''(r)-2rK'(r)+K^2(r)-1=0.
\label{Keqr}
\end{equation}
We can study this further by defining
\begin{equation}
r=e^{-\xi}
\label{xidef}
\end{equation}
so that eq.~(\ref{Keqr}) becomes
\begin{equation}\label{Keq}
\frac{\partial^2 K}{\partial \xi^2} + 3 \frac{\partial K}{\partial \xi}  
- 1 + K^2 =0.
\end{equation}
We may transform this into a more recognisable form by setting
\begin{equation}
\frac{\partial K}{\partial \xi}=-3w(K),
\label{wdef}
\end{equation}
so that eq.~(\ref{Keq}) implies
\begin{equation}\label{Abel2}
w(K) w'(K) -w(K) + \frac{-1 + K^2}{9} = 0.
\end{equation}
This is an Abel equation of the second kind. Unlike some equations of
this type, there appears to be no analytic solution in terms of known
functions. Instead, we may revert back to eq.~(\ref{Keq}) and analyse
it using a method similar to that used by Wu and Yang~\cite{Wu:1967vp}
to construct nonperturbative solutions in pure Yang-Mills
theory. First, one may turn eq.~(\ref{Keq}) into two coupled
first-order differential equations as follows:
\begin{equation}
\left(\frac{\partial K}{\partial \xi},\frac{\partial \psi}{\partial \xi}\right)
=\left(\psi,-3\psi+1-K^2\right).
\label{Keq2}
\end{equation}
This defines a vector field in the $(K,\psi)$ plane, where the curves
that are tangent to this vector field are the solutions we desire. We
show a plot of these curves in figure~\ref{Keq2}, and solutions for
$K(\xi)$ that are finite for all values of $r$ correspond to curves
which are bounded in the plane.
\begin{figure}
\begin{center}
\scalebox{0.45}{\includegraphics{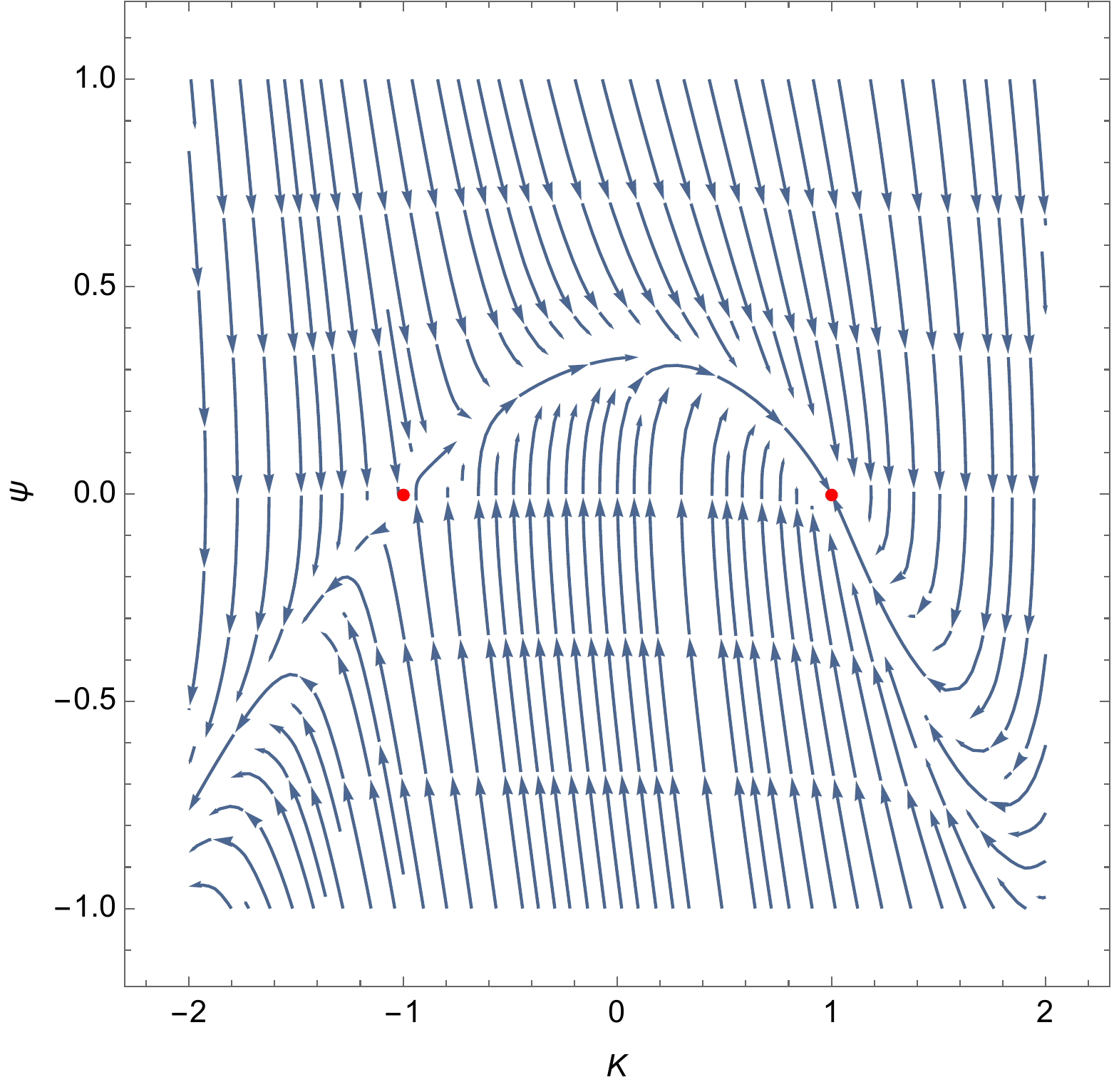}}
\caption{The integral curves of the vector field defined by
  \eqref{Keq2}. The stationary points $(\pm 1,0)$ are marked
  in red.}
\label{fig:curves}
\end{center}
\end{figure}
By inspection, and noting that the vector field is zero at
$(K,\psi)=(\pm 1,0)$, there are three possibilities:
\begin{enumerate}
\item $K(r)=1$: this corresponds to $\bar{S}(r)=0$, and hence the trivial
  solution $\Phi^{aa'}=0$. 
\item $K(r)=-1$: this yields $\bar{S}(r)=-2/r^2$, which is the
  solution already found in ref.~\cite{White:2016jzc}, and reported
  here in eq.~(\ref{phisol1}).
\item $K(r)\rightarrow \pm 1$ as $\xi\rightarrow \pm\infty$
  respectively. This is a non-trivial solution corresponding to the
  single curve that flows from $(-1,0)$ to $(+1,0)$ in
  figure~\ref{fig:curves}. It has not been previously obtained.
\end{enumerate}
Let us analyse the new solution yet further. In fact, this is a
one-parameter family of solutions: if $(K(\xi),\psi(\xi))$ solve
eq.~(\ref{Keq2}), then $(K(\xi-\xi_0),\psi(\xi-\xi_0))$, for arbitrary
constant $\xi_0$, are also solutions. Translating back to the spatial
coordinate $r$, this implies that if $K(r)$ solves eq.~(\ref{Keq}), so
does the rescaled function $K(r/\lambda)$, for general constant
$\lambda$. \\

\begin{figure}
\begin{center}
(a) \scalebox{0.4}{\includegraphics{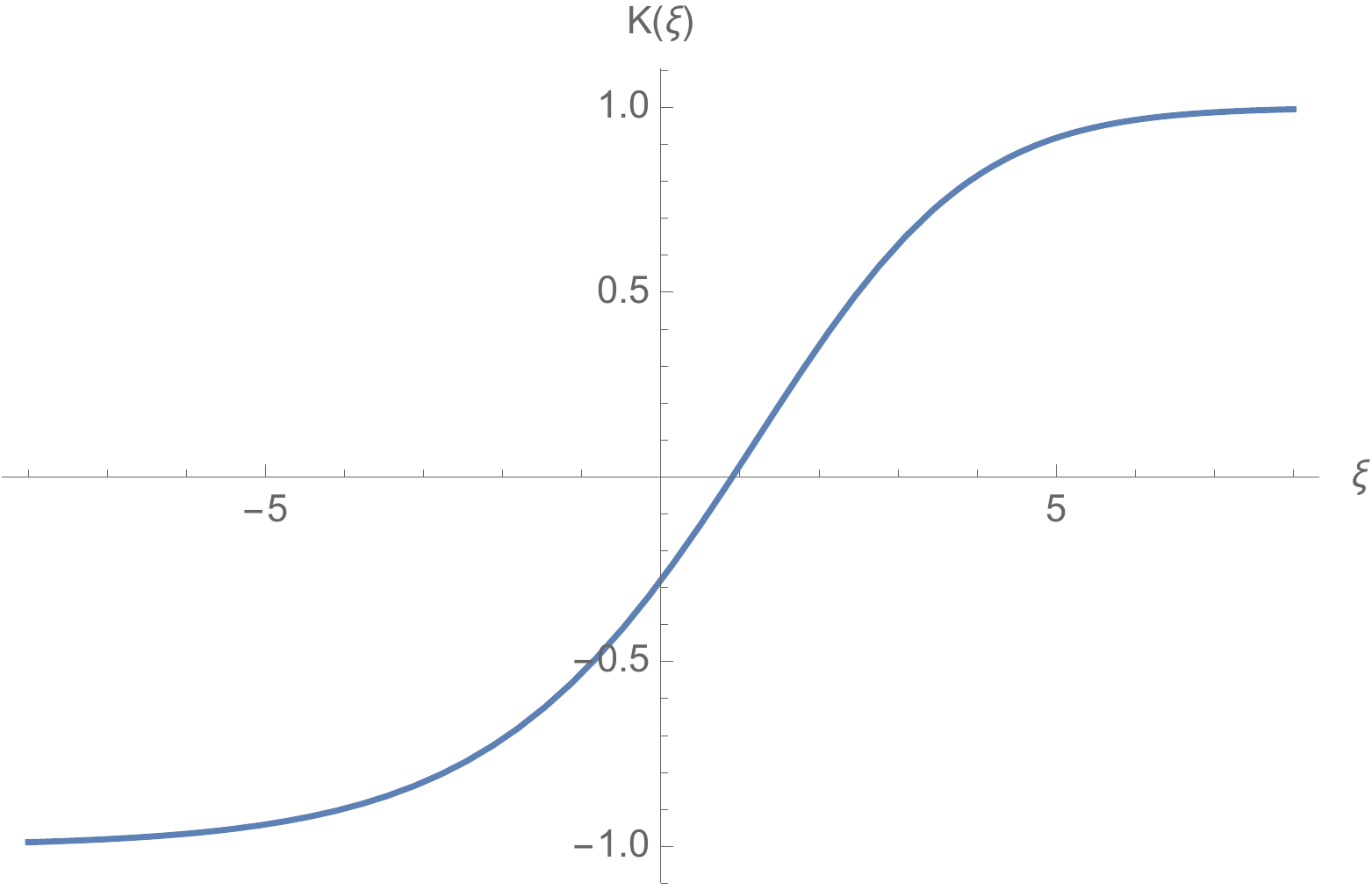}}
\hspace{1cm}
(b) \scalebox{0.4}{\includegraphics{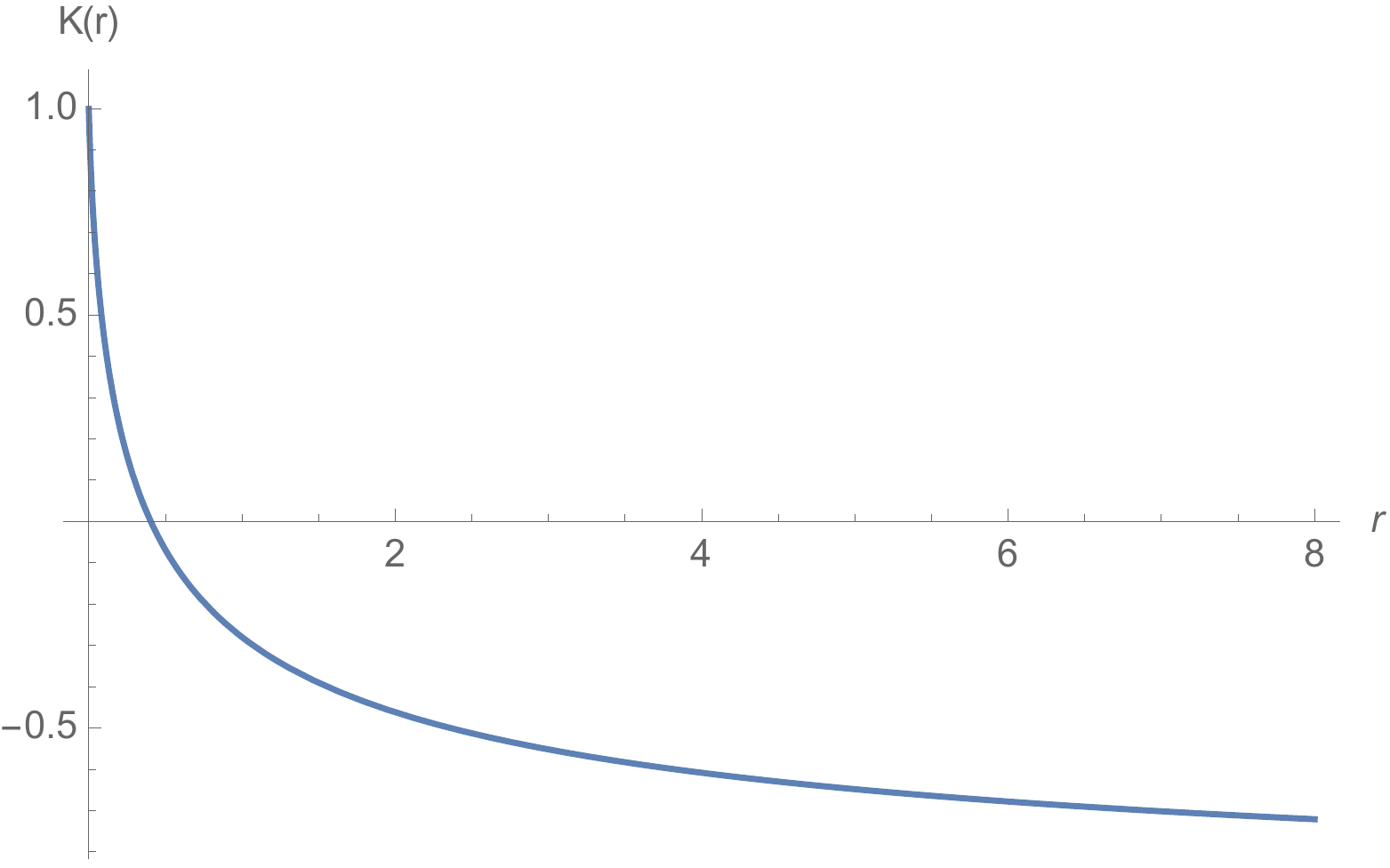}}
\caption{(a) Numerical solution of \eqref{Keq} with asymptotic behaviour
  given by \eqref{Klims}. We have chosen $A=1$; other choices of
  $A$ will lead to the same graph but translated along the $\xi$-axis.
(b) The behaviour of $K$ as a function of $r$.
}
\label{fig:Kgraph}
\end{center}
\end{figure}

Although a complete analytic solution for $K(\xi)$ is not possible, we
can examine its behaviour in asymptotic limits. For
$\xi\rightarrow-\infty$ one may write
\begin{displaymath}
K(\xi)=-1+f(\xi).
\end{displaymath}
Upon substituting this into eq.~(\ref{Keq}) and neglecting terms
quadratic and higher in $f(\xi)$, one finds the general solution
\begin{equation}
f(\xi) =A e^{\left(\frac{\sqrt{17}}{2}-\frac{3}{2}\right) \xi}
+  B e^{\left(-\frac{3}{2}-\frac{\sqrt{17}}{2}\right) \xi}.
\label{ylim1}
\end{equation}
The boundary condition $f(\xi)\rightarrow 0$ as $\xi\rightarrow-\infty$
then requires $B=0$. Similarly, for $\xi\rightarrow+\infty$ we may
substitute
\begin{displaymath}
K(\xi)=1+g(\xi)
\end{displaymath}
in eq.~(\ref{Keq}) and ignore terms quadratic and higher in $g(\xi)$,
leading to the general solution
\begin{equation}
g(\xi)=A'e^{-\xi}+B' e^{-2\xi},
\label{ylim2}
\end{equation}
where the first term gives the dominant behaviour. We conclude that
\begin{equation}
K(\xi)\simeq \begin{cases} 
-1+Ae^{\left(\frac{\sqrt{17}}{2}-\frac32\right)\xi},
&\quad \xi\rightarrow-\infty\\
+1+A'e^{-\xi},&\quad \xi\rightarrow+\infty
\end{cases},
\label{Klims}
\end{equation}
or, translating back to the function $\bar{S}(r)$, 
\begin{equation}
\bar{S}(r)\simeq\begin{cases}
\frac{1}{r^2}\left[
-2+Ar^{-\frac{\sqrt{17}}{2}+\frac32}\right],
&\quad r\rightarrow \infty\\
\frac{A'}{r},&\quad r\rightarrow 0.
\end{cases}
\label{Klims2}
\end{equation}
The two parameters $A$ and $A'$ are not independent. As discussed
above, all solutions for $K(r)$ constitute a rescaling of a canonical
solution. The scaling is fixed by choosing either $A$ or $A'$, such
that the other parameter also becomes fixed. For the particular choice
$A=1$, we show a numerical solution for $K(\xi)$ in
figure~\ref{fig:Kgraph}.\\

The simple power-like solution of eq.~(\ref{phisol1}) is reminiscent
of a Coulomb solution in gauge theory, with a correspondingly
divergent field at the origin~\footnote{The fact that the power
  associated with the divergence is different to the Coulomb solution
  can be traced to the fact that the coupling constant $y$ is
  dimensionful~\cite{White:2016jzc}.}. The new solution found here
consists of dressing the previous solution with a form factor, such
that the divergence at $r\rightarrow 0$ is partially screened. That
is, the $r^{-2}$ behaviour is softened to $r^{-1}$, and by choosing
the rescaling parameter $\lambda$ in $K(r/\lambda)$, this screening
can be made as smooth or abrupt as desired. The softening of the
divergence in the field is also reflected in the energy of the
solution. From eq.~(\ref{Lagrangian}), the Hamiltonian density of the
biadjoint scalar theory is found to be~\cite{White:2016jzc}
\begin{equation}
{\cal H}=\frac12\left[(\dot{\Phi}^{aa'})^2+\nabla\Phi^{aa'}\cdot 
\nabla\Phi^{aa'}\right]-\frac{y}{3}f^{abc}\tilde{f}^{a'b'c'}
\Phi^{aa'}\Phi^{bb'}\Phi^{cc'},
\label{Hamiltonian}
\end{equation}
which for the ansatz of eq.~(\ref{ansatz1}) evaluates to
\begin{equation}
{\cal H} =  \frac{ \cal N}{y^2 T_A^2}
\left( \frac{1}{2} \bar{S}'(r)^2 - \frac{1}{3} \bar{S}(r)^3 \right),
\label{HSbar}
\end{equation}
with ${\cal N}$ the dimension of the (common) Lie group. The energy is
divergent due to the singularity at the origin, but can be evaluated
by implementing a short-distance cutoff $r_0$, corresponding to a
{\it charge radius}:
\begin{equation}
E=4\pi\int_{r_0}^\infty dr\, r^2\,{\cal H}.
\label{Edef}
\end{equation}
For the solution with the asymptotic behaviour of eq.~(\ref{Klims2}),
this gives (as $r_0\rightarrow 0$)
\begin{equation}
E\simeq \frac{\cal N}{y^2\,T_A^2}\frac{2\pi (A')^2}{r_0},
\label{Eres1}
\end{equation}
which is a much softer divergence than the energy associated with
eq.~(\ref{phisol1}), that behaves as $E\sim
r_0^{-3}$~\cite{White:2016jzc}.

\section{Extended solutions for $SU(2)\otimes SU(2)$}
\label{sec:SU(2)}

Having succeeded in constructing an extended solution for the case in
which both gauge groups are identical, let us see if further such
solutions are possible upon restricting each gauge group to be
$SU(2)$, for which one may make a similar ansatz to that used in
ref.~\cite{White:2016jzc}
\begin{equation}
\Phi^{aa'}=\frac{1}{r^2} \left(
A(r)\delta^{aa'}+B(r)\frac{x^a\,x^{a'}}{r^2}
+C(r)\epsilon^{aa'd}\frac{x^d}{r}\right).
\label{Phiansatz}
\end{equation}
Introducing $\bar{A}=A/y$ etc., eq.~(\ref{EOM}) then implies the
coupled non-linear differential equations
\begin{align}
r^2 \bar{A}''-2 r \bar{A}'+2 \bar{A} \bar{B}+2 \bar{A}^2+2
   \bar{A}+2 \bar{B} &=0;\label{A2}\\
   r^2 \bar{B}''-2 r
   \bar{B}'-2 \bar{A} \bar{B}-4 \bar{B}+2 \bar{C}^2&=0;\label{B2}\\
   r^2 \bar{C}''-2 r \bar{C}'+2 \bar{A} \bar{C}+2 \bar{B}
   \bar{C}&=0.\label{C2}
\end{align}
To simplify this we make a further ansatz, namely we write
\begin{align}
\bar{A}(r) &= c_1 f(r) + c_2\\
\bar{B}(r) &= c_3 f(r) + c_4\\
\bar{C}(r) &= c_5 f(r) + c_6,
\end{align}
and then we look for special values of the constants $c_1, \ldots,
c_6$ so that the three equations (\ref{A2} - \ref{C2}) reduce to a
single differential equation for $f(r)$. We thus find two new types of
solution that were not presented previously in
ref.~\cite{White:2016jzc}. 

\subsection{Multiple power-like solution}
\label{sec:power}

The first new solution is obtained upon choosing 
\begin{equation}
\bar{A}(r)=-1,\quad \bar{C}(r)=0,
\label{ACvals}
\end{equation}
in which case $\bar{B}(r)$ satisfies the linear homogeneous
differential equation
\begin{equation}
r^2\bar{B}''(r)-2r\bar{B}'(r)-2\bar{B}(r)=0,
\label{Bdiffeq}
\end{equation}
whose general solution is
\begin{equation}\label{Bsol}
\bar{B}(r) = r^{3/2} \left( b_1\, r^{\frac{\sqrt{17}}{2}} + b_2\,
r^{-\frac{\sqrt{17}}{2}} \right),
\end{equation}
where $b_1$ and $b_2$ are arbitrary constants. The full solution for
$\Phi^{aa'}$ is then
\begin{align}
\Phi^{aa'}&=\frac{1}{yr^2}\left[-\delta^{aa'}+
\left( b_1\, r^{\frac{3}{2}+ \frac{\sqrt{17}}{2}} 
+ b_2\, r^{\frac{3}{2}-\frac{\sqrt{17}}{2}}\right)
\frac{x^a\,x^{a'}}{r^2}
\right]\notag\\
&\simeq \frac{1}{yr^2}\left[-\delta^{aa'}+
\left( b_1\, r^{3.562} 
+ b_2\, r^{-0.562}\right)
\frac{x^a\,x^{a'}}{r^2}
\right].
\label{phisol3}
\end{align}
Here the second term in the square brackets has a part which diverges
more rapidly at the origin than the $r^{-2}$ behaviour of the
solutions of ref.~\cite{White:2016jzc}, and a term which is softer. If
one requires a finite energy upon integrating to infinity, however,
one must set $b_1=0$. The energy, from eq.~(\ref{Edef}), is then found
to be
\begin{equation}
E = 4 \pi  \left(\frac{1}{4} \left(1+\sqrt{17}\right) b_2^2\, r_0^{-\sqrt{17}}-2 b_2\, r_0^{-\frac{3}{2}-\frac{\sqrt{17}}{2}}+\frac{8}{3 r_0^3}\right),
\label{Esol3}
\end{equation}
where again $r_0$ is a small-distance cutoff. For small $r_0$, the
energy diverges as $E \sim r_0^{-\sqrt{17}}\simeq r_0^{-4.123}$
(assuming $b_2 \neq 0$). Interestingly, the solution of
eq.~(\ref{phisol3}) involves the same power of
$r^{\frac32-\frac{\sqrt{17}}{2}}$ as appears in the asymptotic
behaviour for the extended solution of the previous section,
eq.~(\ref{Klims2}).

\subsection{Extended solutions}
\label{sec:extended}

We have also found extended solutions of eqs.~(\ref{A2}--\ref{C2}),
that do not have a pure power-like form. For these solutions, one has
\begin{equation}
\bar{A}(r)=-1+\frac{1}{2}(c^2-1)\bar{B}(r),\quad \bar{C}(r)=c\bar{B}(r),
\label{ABCsol4}
\end{equation}
where $c$ is a constant, and $B(r)$ satisfies
\begin{equation}
r^2\bar{B}''(r)-2r\bar{B}'(r)+(c^2+1)\bar{B}^2(r)-2\bar{B}(r)=0.
\label{Beqsol4}
\end{equation}
Upon writing
\begin{equation}
\bar{B}(r)=\frac{1}{c^2+1}\left(K(r)+1\right),
\label{Brsol4}
\end{equation}
one finds that $K(r)$ satisfies eq.~(\ref{Keqr}). Thus, an extended
solution for $\Phi^{aa'}$ is given by
\begin{equation}
\Phi^{aa'}=\frac{1}{yr^2}
\left[ -\delta^{aa'} + 
\frac{ K(r) + 1}{ c^2  + 1} \left( \frac{c^2-1}{2} \delta^{aa'} 
+\frac{x^a\,x^{a'}}{r^2}+c\,\epsilon^{aa'd}\frac{x^d}{r}\right)\right].
\label{phisol4}
\end{equation}
In section~\ref{sec:common} we found three solutions for $K(r)$, which
lead to the following cases:
\begin{enumerate}
\item $K(r)=1$: in this case, eq.~(\ref{phisol4}) reduces to
\begin{equation}
\Phi^{aa'}=\frac{1}{yr^2}\left[
-\frac{2}{c^2+1}\left(\delta^{aa'}-\frac{x^a\,x^{a'}}{r^2}\right)
+\frac{2c}{c^2+1}\epsilon^{aa'd}\frac{x^d}{r}
\right].
\label{phisol5}
\end{equation}
By replacing
\begin{equation}
c\rightarrow\pm\sqrt{\frac{2-k}{k}},
\label{creplace}
\end{equation}
we see that eq.~(\ref{phisol5}) is the same as the solution of
eq.~(\ref{phisol2}), that was already presented in
ref.~\cite{White:2016jzc}.
\item $K(r)=-1$: in this case, eq.~(\ref{phisol4}) reduces to
\begin{displaymath}
\Phi^{aa'}=-\frac{1}{yr^2}\delta^{aa'},
\end{displaymath}
which is the same as eq.~(\ref{phisol1}) for the special case in which
both gauge groups are $SU(2)$, so that $T_A=2$.
\item The most general solution has $K(r)$ given by the function of
  figure~\ref{fig:Kgraph}, with asymptotic limits given by
  eq.~(\ref{Klims2}). This solution is new.
\end{enumerate}
The extended solution no longer has the pure-axial property of
eq.~(\ref{phisol2}), consistent with the fact noted in
ref.~\cite{White:2016jzc} that extended axial solutions cannot
exist. The energy of the solution in case 3 above, in terms of the
usual short-distance cutoff $r_0$, is found to have the leading
behaviour (as $r_0\rightarrow 0$)
\begin{equation}
E\simeq\frac{32\pi}{y^2}\frac{1}{(1+c^2)r_0^3}.
\label{Esol4}
\end{equation}
Note that the extended solution of section~\ref{sec:common} (where
again $T_A=2$) can also be obtained from eq.~(\ref{phisol4}), by
choosing $c\rightarrow\infty$:
\begin{equation}
\Phi^{aa'}\xrightarrow{c\rightarrow\infty}
\frac{K(r)-1}{2yr^2}\delta^{aa'}.
\label{phisol4lim}
\end{equation}
Finally, we note that a complex solution is also possible. By choosing
$c^2=-1$ in eq.~(\ref{Beqsol4}), this reduces to eq.~(\ref{Bdiffeq}),
whose general solution is given by eq.~(\ref{Bsol}). One then obtains
\begin{equation}
\Phi^{aa'}=\frac{1}{yr^2}\left[
-\delta^{aa'}+b_2 r^{\frac32-\frac{\sqrt{17}}{2}}
\left(-\delta^{aa'}+\frac{x^a\,x^{a'}}{r^2}
\pm i \epsilon^{aa'd}\frac{x^d}{r}\right)
\right],
\label{phisol6}
\end{equation}
where we have again set $b_1\rightarrow 0$ to ensure that the energy
is bounded at spatial infinity.

\section{Conclusion}
\label{sec:conclude}

Biadjoint scalar field theory occurs in a number of contexts, and
plays an intriguing role in determining the dynamics of perturbative
scattering amplitudes in gauge and gravity theories. The
nonperturbative properties of this theory remain relatively
unexplored, and in this paper we have presented a number of new
solutions involving inverse powers of the coupling constant. Unlike
the first solutions presented in ref.~\cite{White:2016jzc}, the
results of the present paper have an extended structure, indicating
that the spectrum of nonperturbative solutions of biadjoint scalar
theory is much richer than has previously been
suggested. Interestingly, the solutions presented here include those
in which the strong divergent behaviour of the field at the origin is
partially screened. \\

There are a number of avenues for further work. Firstly, it may be
possible to find more solutions of the biadjoint scalar theory,
including non-static solutions. Secondly, one may imagine coupling the
biadjoint scalar to a gauge field, as occurs in some applications of
the double copy. Finally, the question of whether the nonperturbative
solutions found here and in ref.~\cite{White:2016jzc} can themselves
be copied to gauge theory or gravity deserves further attention.


\section*{Acknowledgments}

CDW is supported by the UK Science and Technology Facilities Council
(STFC).

\bibliography{refs.bib}
\end{document}